\documentclass[journal=jpccck,manuscript=article]{achemso}

\usepackage[version=3]{mhchem} 
\usepackage[T1]{fontenc}       
\usepackage{amssymb}
\usepackage{tabularx,ragged2e}
\newcommand{\onlinecite}[1]{\hspace{-1 ex} \nocite{#1}\citenum{#1}}

\SectionNumbersOn

\usepackage{color, hyperref, multirow}

\author{Huan Tran}
\affiliation{School of Materials Science and Engineering, Georgia Institute of Technology, 771 Ferst Drive NW, Atlanta, GA 30332, USA}
\author{Kuan-Hsuan Shen}
\affiliation{School of Materials Science and Engineering, Georgia Institute of Technology, 771 Ferst Drive NW, Atlanta, GA 30332, USA}
\author{Shivank Shukla}
\affiliation{School of Materials Science and Engineering, Georgia Institute of Technology, 771 Ferst Drive NW, Atlanta, GA 30332, USA}
\author{Ha-Kyung Kwon}
\affiliation{Energy and Materials, Toyota Research Institute, Los Altos, California 94022, USA}
\author{Rampi Ramprasad}
\email{rampi.ramprasad@mse.gatech.edu}
\affiliation{School of Materials Science and Engineering, Georgia Institute of Technology, 771 Ferst Drive NW, Atlanta, GA 30332, USA}

\title{Informatics-Driven Selection of Polymers for Fuel-Cell Applications}

\begin{document}



\begin{abstract}
Modern fuel cell technologies use Nafion as the material of choice for the proton exchange membrane (PEM) and as the binding material (ionomer), used to assemble the catalyst layers of the anode and cathode. These applications demand high proton conductivity as well as other requirements. For example, PEM is expected to block electrons, oxygen, and hydrogen from penetrating and diffusing while the anode/cathode ionomer should allow hydrogen/oxygen to move easily, so that they can reach the catalyst nanoparticles. Given some of the well-known limits of Nafion, such as low glass-transition temperature, the community is in the midst of an active search for Nafion replacements. In this work, we present an informatics-based scheme to search large polymer chemical spaces, which includes establishing a list of properties needed for the targeted applications, developing predictive machine-learning models for these properties, defining a search space, and using the developed models to screen the search space. Using the scheme, we have identified 60 new polymer candidates for PEM, anode ionomer, and cathode ionomer that we hope will be advanced to the next step, i.e., validating the designs through synthesis and testing. The proposed informatics scheme is generic, and can be used to select polymers for multiple applications in the future.

\end{abstract}
\newpage
\section{Introduction}

The history of fuel cells, the devices that directly convert the chemical energy from reactants such as hydrogen (fuel) and oxygen (oxidant) into electricity, dates back to the 1840s.\cite{groove1842on, hoogers2002fuel, li2005principles} Since then, significant efforts and progress have been made to advance fuel cells so that they can find applications in transportation\cite{yoshida2015toyota,pollet2019current} and other sectors, e.g., consumer electronics, residential power supply, and back-up power for banks and telecommunication companies.\cite{hickner2005chemical, wang2020materials, kusoglu2017new, zhu2022recent}. As the only byproducts of fuel cells are water and waste heat, such devices are ideal power generators and a particularly useful energy conversion method for a future clean economy. A typical fuel cell is composed of three active components, i.e., a fuel electrode (anode), an oxidant electrode (cathode), and an electrolyte filled inbetween. The electrolyte is a material that allows positive ions, e.g., protons, to transport easily while blocking electrons and the reactants, e.g., hydrogen and oxygen gases, from penetrating and diffusing. Fuel cells are classified based on the electrolyte material used, e.g., alkaline fuel cells, phosphoric acid fuel cells, molten carbonate fuel cells, solid oxide fuel cells, and proton/anion exchange membrane (PEM/AEM) fuel cells.\cite{grubb1959patent,grubb1960batteries, hoogers2002fuel, li2005principles, zhang2012recent,wang2020materials, kusoglu2017new, zhu2022recent, ramaswamy2019alkaline,zou2021machine}

\begin{figure}[t]
\centering
\includegraphics[width=0.8\linewidth]{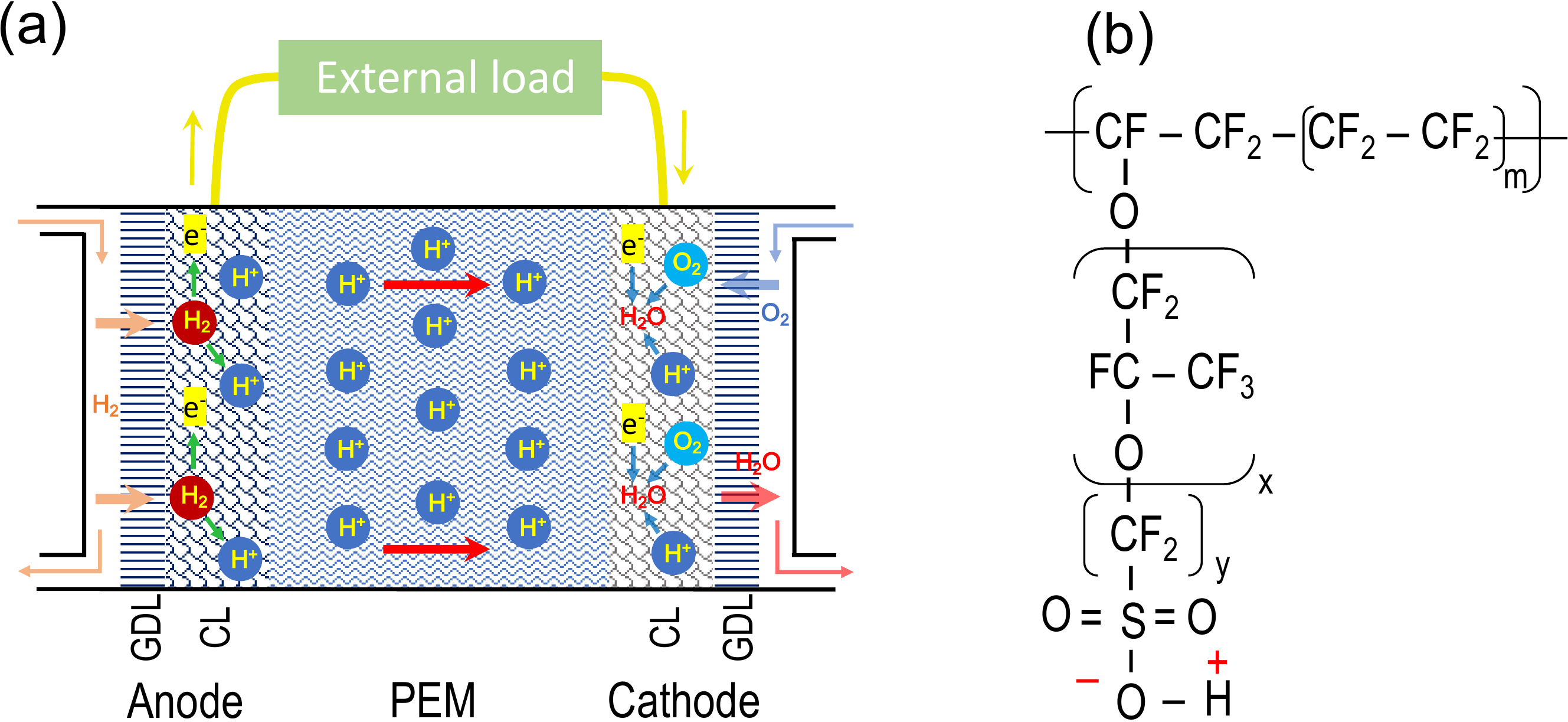}
\caption{(a) Schematic of a cross-sectional view of a PEM fuel cell unit in which a PEM layer is sandwiched between two electrodes (anode and cathode), each of them contains a gas diffusion layer (GDL) and a catalyst layer (CL), and (b) general chemical structure of Nafion, a random copolymer composed of an electrically neutral semicrystalline polytetraﬂuoro-ethylene backbone and a randomly tethered side-chain ending with the pendant sulfonate group -SO$_3^-$. The backbone length of Nafion is $m \simeq 5.5$ while the polar, hydrophilic -SO$_3^-$ sulfonate groups are essential for capturing water molecules. The catalyst layers are created by using polymeric ionomers to bind Pt nanoparticles together.}\label{fig:fuelcell}
\end{figure}

In the current PEM fuel cell technology, perfluorosulfonic acid (PFSA) polymers \cite{grubb1959patent,grubb1960batteries} are used not only as the (electrolyte) PEM \cite{zhang2012recent} but also as the binding material, i.e., ionomers, in the catalyst layers of the anode and cathode,\cite{suter2021engineering, berlinger2021multicomponent, zhang2008pem,tanaka2022aromatic} as schematically shown in Fig. \ref{fig:fuelcell} (a). Technically, each electrode contains a gas diffusion layer and a catalyst layer, prepared by using a solution of a polymeric ionomer, e.g., Teflon or Nafion, to bind nanoparticles of electrocatalysts, e.g., Pt, together and to a support.\cite{holdcroft2014fuel, jinnouchi2021role, zhao19} At this time, designs of catalyst layers without ionomers remain impractical because of their extremely low durability.\cite{rabat2008plasma, saha2006high} During operation, hydrogen is split into electrons and protons (the hydrogen oxidation reactions) in the anode catalyst layer. The electrons flow through the outer circuit and the protons are transported through the PEM layer before combining with oxygen to form water (the oxygen reduction reactions) in the cathode catalyst layer, thus closing the circuit. For state-of-the-art PEM fuel cells that use H$_2$/air and layers of high surface area carbon-supported Pt/Pt-alloy-based catalysts, the power density could reach 900--1000 mW/cm$^2$ at cell voltages $\leq 0.65$ V (80 $^\circ$C, 100\% relative humidity RH, and 150 kPa outlet pressure).\cite{spendelow2011progress,ramaswamy2019alkaline}

Nafion, whose chemical structure is shown in Fig. \ref{fig:fuelcell} (b) and key properties summarized in Table \ref{table:nafion}, is a PFSA polymer that is used predominantly as the PEM in fuel cells. Although the proton conductivity of Nafion is high, its very strong dependence on the amount of water in the membrane, as discussed in Sec. \ref{sec:data}, generates significant challenges in humidification and water management.\cite{jiao2011water,kusoglu2017new,zhu2022recent,chang2018humidification} While Nafion is used as the cathode ionomer, its low oxygen permeability \cite{kudo2016humidity,kudo2013analysis} has motivated numerous works to search for new binding materials.\cite{jinnouchi2021role,tanaka2022aromatic,suzuki2011ionomer} In fact, the catalyst layers of a fuel cell are necessarily complex, generating a highly active research area\cite{suter2021engineering, berlinger2021multicomponent, gasteiger2004dependence} aimed at improving the reaction rate and reducing the loading of Pt, a precious metal. Moreover, the fully flurionated chemical structure of Nafion also leads to high production cost while its relatively low glass transition temperature ($T_{\rm g} \simeq 120^\circ$C)\cite{jung2012role, meyer2017investigation} limits the fuel cell operation at high temperature conditions. Previous attempts to develop alternatives to Nafion have not been very successful, with no comparable material identified and optimized so far.\cite{miyake2017design,souzy2005proton, wang2020fundamentals,kraytsberg2014review} 

\begin{figure}[t]
\centering
\includegraphics[width=0.5\linewidth]{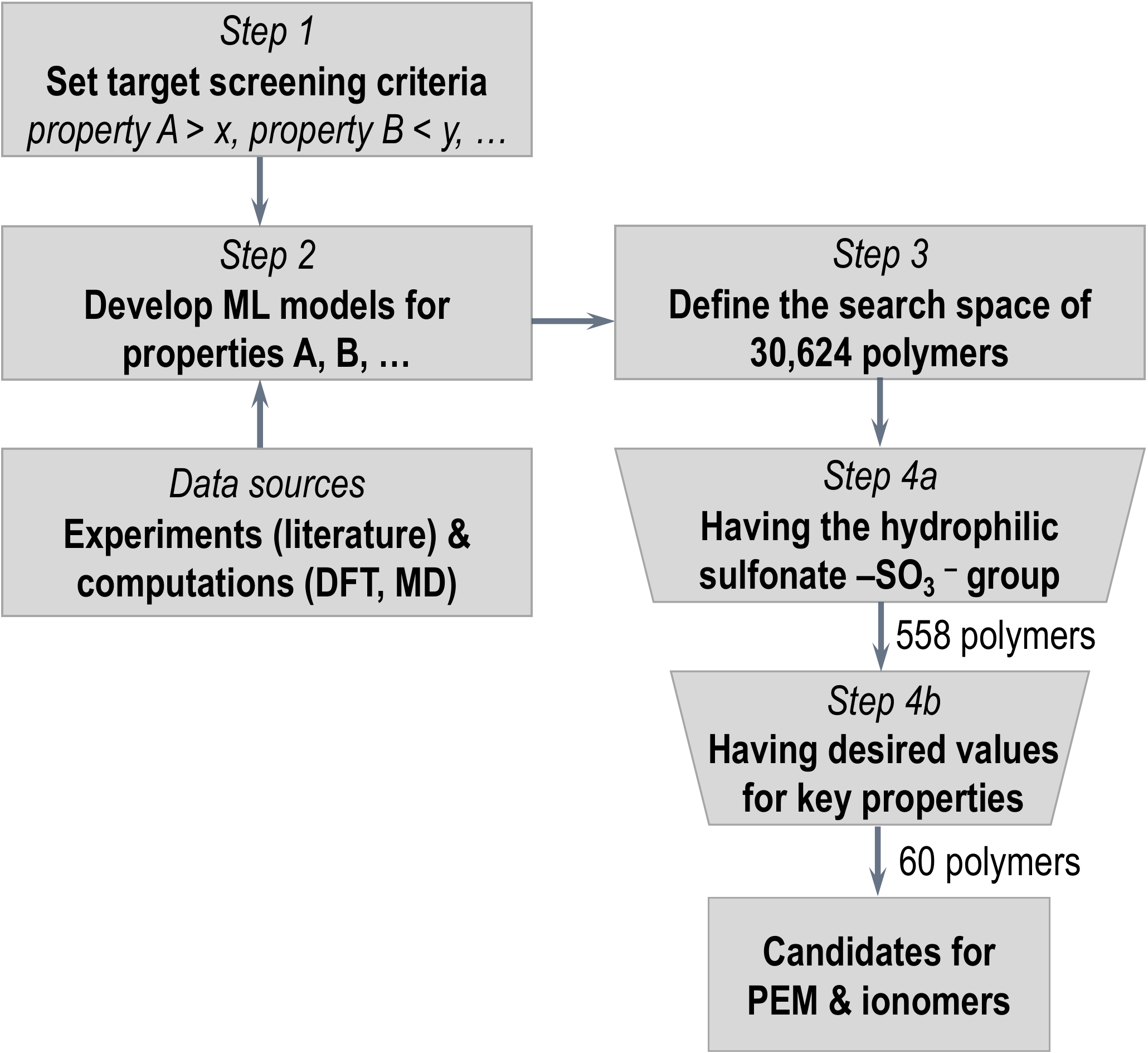}
	\caption{A machine-learning and multi-objective driven scheme for the selection of polymers that can be used as a PEM or (anode/cathode) ionomer in fuel cells.}\label{fig:scheme}
\end{figure}

The goal of this work is to identify good alternatives for PEM and ionomers from a vast space of known polymers using an artificial intelligence (AI)-based screening strategy, outlined in Fig. \ref{fig:scheme}. This search benefits from recent advancements within the domain of polymer informatics.\cite{Rampi:ML, chen2021polymer, kuenneth2021polymer, Chiho:PG, doan2020machine, Huan:design, Arun:design, Chiho:GA, Rohit:VAE,gurnani2021polyg2g, kamal2021novel,kamal2020computable} In the first step, target screening criteria, given in terms of desired properties are identified. Then, machine-learning (ML) models for quickly predicting these properties are developed and validated, using experimental and/or computational data. Next, a search space is defined in step 3, creating a big candidate set whose key properties are unknown. In step 4, ML models are used to predict the key properties of the candidates, and those that pass the screening criteria are identified. The final step is devoted to validations using computations and/or experiments. We note that when the search space is large, i.e., practically infinite, advanced methods like genetic algorithm\cite{Chiho:GA}, variational auto-encoder\cite{Rohit:VAE}, and polyG2G\cite{gurnani2021polyg2g} can be used for more efficient searches for materials with targeted properties by inverting the property prediction pipeline. 

In the present work, we have examined and established a list of polymer properties that are needed for PEM, anode ionomer, and cathode ionomer. Predictive models were developed for these properties, validated, and used to screen polymers spanning a search space of 30,624 known polymers. A total of 48 new candidates for PEM, 10 new candidates for anode ionomer, and 2 new candidates for cathode ionomer have been identified. All of them are predicted to have superior proton conductivity, the most important property needed for PEM and ionomers. We anticipate that the candidates reported in this work will be advanced to the step of experimental synthesis and testing in the near future.

\section{Methods}

\subsection{Key properties and search criteria}
Proton conductivity $\sigma$ is the most important property for fuel cell PEM and ionomers.\cite{kusoglu2017new,hickner2005chemical,zhu2022recent,suter2021engineering, berlinger2021multicomponent} Two main mechanisms of proton conduction in fuel cell PEM are vehicular, i.e., protons are bound to and carried by the water molecules when they diffuse, and hopping, i.e., protons hop from one water molecule to another. In other words, proton transport relies heavily on the water molecules contained in a network of connected water channels in the materials.\cite{li2005principles,jiao2011water,kusoglu2017new, kusoglu2012water, hickner2005chemical} In the channels, typically formed due to phase separation, the polar, hydrophilic pendant sulfonate groups -SO$_3^-$ (see Fig. \ref{fig:fuelcell} (b)) are well-known\cite{lowry1980investigation,cable1995effects, korzeniewski2008responses,kusoglu2017new,souzy2005proton,liu2022comb} to be critical for capturing and holding the water molecules. For this reason, the proton conductivity $\sigma$ and the water uptake $\lambda$ are very closely related properties, and they are usually reported simultaneously. When $\sigma$ is measured experimentally, samples are subjected to a given temperature, e.g., $T=80^\circ$C at working condition, and a method of humidifying, i.e., if the samples are submerged in water vapor or liquid water. If water vapor is used, $\sigma$ is typically measured for a range of relative humidity RH $\simeq 10-100$\% (see Fig. \ref{fig:models} (b) for some experimental data) while if liquid water is used, RH is 100\% and only a value of $\sigma$ is given. 

Gas permeability is the next important requirement for the materials used for fuel cell PEM and ionomers, which also depends on the temperature and relative humidity.\cite{suter2021engineering, berlinger2021multicomponent, jinnouchi2021role, park2019does} Because the PEM layer must block hydrogen and oxygen gases from penetrating and diffusing, its oxygen and hydrogen permeabilities, i.e., $\mu_{\rm O_2}$ and $\mu_{\rm H_2}$, must be low. On the other hand, the cathode ionomer should allow oxygen gas to flow well to reach the catalyst nanoparticles, thus a high $\mu_{\rm O_2}$ is desirable while the value of $\mu_{\rm H_2}$ is not critical because the hydrogen gas is assumed to be blocked by the PEM layer. Likewise, the anode ionomer should have a high $\mu_{\rm H_2}$ without specific requirements on $\mu_{\rm O_2}$. 

Next, the PEM layer in a fuel cell must be a good barrier for electrons, forcing them to run through the external load. This requirement can be translated into a large enough electronic ``band gap'' $E_{\rm g}$ of the PEM material. In addition, a high value of $E_{\rm g}$ is also needed to secure the electrochemical stability of the PEM. As the PEM (electrolyte) layer is sandwiched between two electrodes, the anode electrochemical potential must be lower than the reduction potential, i.e., the conduction band minimum, while the cathode electrochemical potential must be higher than the oxidation potential, i.e., the valence band maximum, of the PEM material.\cite{goodenough2011challenges, goodenough2012rechargeable, chen2019electrochemical,marchiori2020understanding} Given that the experimentally measured open circuit and full load voltages of a PEM fuel cell are $\simeq 1.0$ and $\simeq 0.6$ V, respectively, \cite{cai19,zhang2006pem, shimada2018key} the electronic band gap $E_{\rm g}$ of PEM should be much larger than 1.0 eV. On the other hand, although the catalyst layers should be good electrical conductors, they conduct electrons through the catalyst and support nanoparticles. Therefore, there are no requirements on $E_{\rm g}$ for the anode and cathode inonomers. 

Finally, materials used for PEM and ionomer should also be mechanically and thermally robust. The mechanical robustness of a proton conducting material is important because the water channels through which protons are transported must withstand the internal pressure. These requirements are translated into a list of other desired properties, including high thermal decomposition temperature $T_{\rm d}$, high glass transition temperature $T_{\rm g}$, and high Young's modulus $E$. The properties discussed above for the case of Nafion are summarized in Table \ref{table:nafion}.

\begin{table*}[t]
\caption{Key properties of Nafion for fuel cell applications, roughly given in the descending ordering of the criticality. In most of the cases, experimental values reported in certain ranges, depending on the experimental conditions.}\label{table:nafion}

\begin{tabularx}{\textwidth}{@{} 
	>{\RaggedRight} p{5.0cm}  
	>{\centering} p{1.5cm}  
	>{\centering} p{2.1cm} 
	>{\centering} p{4.05cm} 
	>{\RaggedRight} p{1.65cm} 
	>{\centering} p{0.2cm} 
	}
\hline
\hline
	\multirow{2}{*}{Key property} & \multirow{2}{*}{Unit} & \multicolumn{3}{c}{Experimental data} & \\
\cline{3-5}
	& &Value & Details & Refs. & \\
\hline
	Proton conductivity $\sigma$ &S/cm & $10^{-3} - 10^{-1}$ & $80^\circ$C, 10-100\% RH & \onlinecite{feng2018characterization, zhang2018sulfonated,peng2017preparation, si2012synthesis,wang2012clustered} & \\
	O$_2$ permeability $\mu_{\rm O_2}$ &Barrer &$1.1 - 34.3$& $30-80^\circ$C, 10-90\% RH& \onlinecite{zhang2018sulfonated, mukaddam2016gas, fan2014role, chiou1988gas}&\\
	H$_2$ permeability $\mu_{\rm H_2}$ &Barrer &$9.3 - 65.0$& $30-80^\circ$C, 10-90\% RH& \onlinecite{zhang2018sulfonated, mukaddam2016gas, fan2014role, chiou1988gas, goo2022polyamide}&\\
	Electronic band gap $E_{\rm g}$ &eV & $7.54$ & Computed, DFT & \onlinecite{Huan:Data}& \\
	Glass trans. temp. $T_{\rm g}$ &K & $396-398$ & & \onlinecite{jung2012role, meyer2017investigation}& \\
	Thermal decom. temp. $T_{\rm d}$ &K & $553$ & & \onlinecite{lage2004thermal, samms1996thermal}&\\
	Young's modulus $E$ &MPa & $50-220$ &$30-60^\circ$C, 10-90\% RH & \onlinecite{zhang2018sulfonated, roberti2010measurement, caire2016accelerated}& \\
\hline
\hline
\end{tabularx}
\end{table*}

Our objective here is to discover polymers that can potentially be better than Nafion in three fuel cell applications, i.e., PEM, anode ionomer, and cathode ionomer. The desired property values (i.e., the search criteria) for these applications are compiled in Table \ref{table:criteria}. In particular, selected candidates should have higher predicted proton conductivity than Nafion at RH = 100\% (for liquid water condition) or in the entire range of relative humidity (for water vapor condition). For H$_2$ and O$_2$ permeabilities, the thresholds were selected as the midpoints of the reported value ranges shown in Table \ref{table:nafion}, which are close to the predictions of our ML models, summarized in Table \ref{table:model}. The electronic band gap $E_{\rm g}$ of the polymer (infinite) chain model of Nafion,\cite{Huan:Data, Huan:PSP, sahu2022polymer} computed using density functional theory (DFT),\cite{DFT1, DFT2, vasp1, vasp3} and the Heyd-Scuseria-Ernzerhof exchange-correlation functional\cite{HSE} is $7.54$ eV. This value is close to the upper limit ($\simeq 9.7$ eV) of all the current datasets of polymer band gap $E_{\rm g}$, including the one that is used for this work (see Supporting Information for more details). Given that the band gap $E_{\rm g}$ of a PEM should be much larger than $1.0$ eV (the open-circuit voltage of a PEM fuel cell) while leaving enough room for the PEM candidate selections, a threshold of $4.0$ eV was used. We believe that polymeric materials with such a band gap are already good insulators and electrochemically stable. Finally, because the vast majority (98.5\%) of our proton conductivity data (see Sec. \ref{sec:data}) involve the sulfonate group (-SO$_3^-$), only polymer candidates having this functional group are considered. Given the defined criteria, a candidate for PEM is not suitable for anode and cathode ionomers.

\begin{table*}[t]
\caption{Requirements for key properties of the polymers that can be used as PEM, anode ionomer (AI), and cathode ionomers (CI) in fuel cell.}\label{table:criteria}
\begin{tabularx}{\textwidth}{@{} 
	>{\RaggedRight} p{0.5cm} 
	>{\RaggedRight} p{6.5cm}  
	>{\centering} p{1.5cm} 
	>{\centering} p{1.5cm} 
	>{\centering} p{1.5cm} 
	>{\centering} p{1.5cm} 
	>{\RaggedRight} p{0.1cm}
	}
\hline
\hline
	\multirow{2}{*}{No.}	&\multirow{2}{*}{Key property} & \multirow{2}{*}{Unit} & \multicolumn{3}{c}{Desired value}  \\
\cline{4-6}
	  &  &  & PEM & CI   & AI &\\
\hline
	1	&       Proton conductivity $\sigma$ & S/cm    & High & High   & High &\\
	2	&	O$_2$ permeability $\mu_{\rm O_2}$ & Barrer& $< 18$ & $> 18$ & N/A &\\
	3	&	H$_2$ permeability $\mu_{\rm H_2}$ & Barrer & $< 37$ & N/A & $> 37$ &\\
	4	&	Electronic band gap $E_{\rm g}$ &eV & $> 4.0$ & N/A & N/A &\\
	5	&	Glass trans. temp. $T_{\rm g}$ & K & $>396$ & $> 396$ & $> 396$ &\\
	6	&	Thermal decom. temp. $T_{\rm d}$ &K & $> 573$ & $> 573$ & $>573$ &\\
	7	&	Young's modulus $E$ & MPa & $> 156$ & $> 156$ & $>156$ &\\
	8	&	Having the sulfonate group (-SO$_3^-$) && Yes & Yes & Yes &\\
\hline
\hline
\end{tabularx}
\end{table*}

Many of the key property requirements shown in Table \ref{table:criteria} are conflicting, exposing a common challenge of material discoveries for specific applications. As an example, protons are transported through connected water channels in PEM and ionomers, so polymers that are suitable for this functionality must include either a sufficient fraction of rigid (double and/or triple) bonds or, like Nafion, a combination of hydrophobic backbone and hydrophilic functional groups. Polymers in the first class cannot have high band gap $E_{\rm g}$ while the second class, combining with the requirement of having sulfonate (-SO$_3^-$) group, points directly and only to the variants of Nafion, thus seriously limiting the chemical space exploration. Although this list may further be expanded or modified, we believe that these material property requirements are essential for good candidates of PEM and ionomers in the fuel cell technology.

\subsection{Data and data representation}\label{sec:data}
To search chemical space efficiently, the evaluations of the key properties (shown in Table \ref{table:criteria}) must be rapid. The best answer for this requirement is a series of ML models. During the last decade, the development and applications of ML models for polymer property predictions and rational design have been steadily demonstrated while works in this area continue to mount.\cite{Rampi:ML, chen2021polymer, kuenneth2021polymer, Chiho:PG, doan2020machine, Huan:design, Arun:design, Chiho:GA, Rohit:VAE,gurnani2021polyg2g} In a typical and established ML workflow, polymer data involving targeted properties are generated/curated, numerically represented (fingerprinted), and learned to develop predictive models. \cite{Chiho:PG, doan2020machine, kuenneth2021polymer,Huan:design, Arun:design} These models are then used to rapidly search over vast polymer spaces, discovering those with desired properties for certain applications.\cite{Huan:design, Arun:design, Chiho:GA, Rohit:VAE,gurnani2021polyg2g}

A ML model for proton conductivity is essential for our objective. Because the proton conductivity $\sigma$ and the water uptake $\lambda$ are very closely related properties, we curated two datasets of $\sigma$ and $\lambda$ in polymers, the former containing 2,137 points and the latter containing 1,879 points. About 98.5\% (2,105 entries) of the proton conductivity and 42.6\% (801 entries) of the water uptake dataset involve the ionic -SO$_3^-$ sulfonate group. From the data learning standpoint, datasets of correlated properties can be fused and learned simultaneously  in a {\it multi-task} (MT) ML model so that possible hidden correlations among them can be accessed.\cite{kuenneth2021polymer, Tuoc:PrNN} Likewise, a dataset of 2,624 data points for the permeabilities of six gases, including H$_2$, O$_2$, He, CO$_2$, N$_2$, and CH$_4$, were taken from previous works,\cite{zhu2020polymer, kuenneth2022bioplastic} augmented, and learned in another MT model. Four other datasets of polymer band gap $E_{\rm g}$, thermal decomposition temperature $T_{\rm d}$, glass transition temperature $T_{\rm g}$, and Young's modulus $E$ were also utilized from past works.\cite{Chiho:PG, doan2020machine, kuenneth2022bioplastic} We note that among these datasets, only those for $\sigma$ and $\lambda$ contain the temperature $T$, the relative humidity RH, and if the water is in the liquid form or not, when the measurements were made. Such information is unavailable in the other datasets. These datasets are summarized in Table \ref{table:model}.

Except for the $E_{\rm g}$ dataset that was prepared computationally at the DFT level for 4,121 homopolymers,\cite{Huan:Data, kamal2021novel,kamal2020computable} the other datasets contain experimentally measured data of both homopolymers and copolymers, whose chemical structures are represented using {\sc smiles} strings, which stand for {\it simplified molecular-input line-entry system}.\cite{smiles} Because {\sc smiles} was initially defined for molecules, this concept has been adapted to represent the homopolymer repeat unit by explicitly specifying the connecting points.\cite{Chiho:PG, doan2020machine} In case of copolymers, the {\sc smiles} and the concentration of each component are available while their nature, i.e., if they are block copolymers or random copolymers, are generally unavailable. 

Polymers data, as materials data in general, must be represented numerically in a proper way, or fingerprinted, so that they can be readable by ML algorithms. In this work, we used the fingerprinting scheme developed previously \cite{Pilania_SR, Huan:design, Arun:design, Chiho:PG, Chiho:PG, doan2020machine, kuenneth2021polymer} Within this scheme, the polymer {\sc smiles} representation is converted to $\simeq 3000$ numerical fingerprint components, which are arranged into three categories that correspond to three different length scales, i.e., atomic, block, and chain levels. For copolymers with multiple repeat units and concentrations, their fingerprints are defined as the composition-weighted sum of the fingerprint of all the repeat units.\cite{kuenneth2021polymer} Only in the particular cases of proton conductivity and water uptake, some additional fingerprint components are the temperature $T$, the relative humidity RH, and the measurement condition, i.e., if the samples are submerged in liquid water or not. 

\subsection{Machine learning approach}
For ML algorithm, we used Gaussian process regression (GPR)\cite{GPRBook,GPR95} with a radial basic function kernel. The choice of GPR was motivated by several reasons. First, GPR is explicitly similarity-based and therefore, intuitive. Second, by assuming the output is a realization of a Gaussian process, GPR provides a built-in measure of the prediction uncertainty. Finally, the datasets used in this work are not too big, thus training a GPR model and using it to make predictions is not computationally intensive while overfitting can be better controlled by the internal cross-validation step. 

\subsection{Screening space}
The screening space contains 30,624 known polymers, including 16,858 copolymers and 13,766 homopolymers, all of them have been synthesized and reported in the literature. Some more information on this dataset can be found in Refs. \onlinecite{doan2020machine, kuenneth2021polymer, kuenneth2022bioplastic}. In addition to the {\sc smiles} strings and respective concentrations, references pointing to the report of each polymer are also available. Starting from the {\sc smiles} and corresponding concentration, the fingerprints of all the polymers in the screening space were computed and used as the input for the ML models developed. For proton conductivity predictions, we assumed $T = 80^\circ$C while considering the whole range of relative humidity RH and two humidifying methods, e.g., the sample is submerged in liquid water or water vapor. Those satisfying the criteria for PEM, anode ionomers, and cathode ionomers as summarized in Table \ref{table:criteria} are compiled in three list of candidates for further considerations, perhaps experimentally.

\section{Results}
\subsection{ML models}
The first multi-task ML model, i.e., M1, was trained on two datasets of $\sigma$ and $\lambda$ for predicting $\sigma$. Likewise, M2 is the second MT model that was trained on six datasets of O$_2$, H$_2$, N$_2$, He, CO$_2$, and CH$_4$ permeabilities for predicting $\mu_{\rm O_2}$ and $\mu_{\rm H_2}$. Technically, the dataset used for each multi-task model was prepared by merging the datasets of $n$ correlated properties ($n = 2$ for M1 and $n=6$ for M2), each property is specified by an additional $n-$dimension vector stacked to the fingerprint computed from the {\sc smiles}. This data-fusion technique has been demonstrated in exploiting the possible hidden correlations among the related materials properties.\cite{kuenneth2021polymer,zhu2020polymer,Tuoc:PrNN} For the other four key properties, i.e., $E_{\rm g}$, $T_{\rm d}$, $T_{\rm g}$, and $E$, four ML models, i.e., M3, M4, M5, and M6, were developed. Given that the training data of M1 and M2 span over 10 orders of magnitude, these models were trained in the log scale and the predictions were then transformed back to the real (linear) scale. Moreover, a dimensionless error metric named {\it mean order of magnitude error} (MOME), defined as $\langle{\rm abs}(\log_{10}(p_{\rm pred}/p_{\rm ref}))\rangle$, where $p_{\rm pred}$ and $p_{\rm ref}$ are the predicted and reference values and $\langle\cdots\rangle$ stand for the average over the predictions, was used to evaluate M1 and M2. On the other hand, M3, M4, M5, and M6 were trained on the linear scales and root-mean-square error (RMSE) is a suitable error metric for them.  Table \ref{table:model} provides a summary of these models and the datasets they were trained on while the visualized performances of these models can be found in Supporting Information. 

\begin{figure}[t]
\centering
\includegraphics[width=1.0\linewidth]{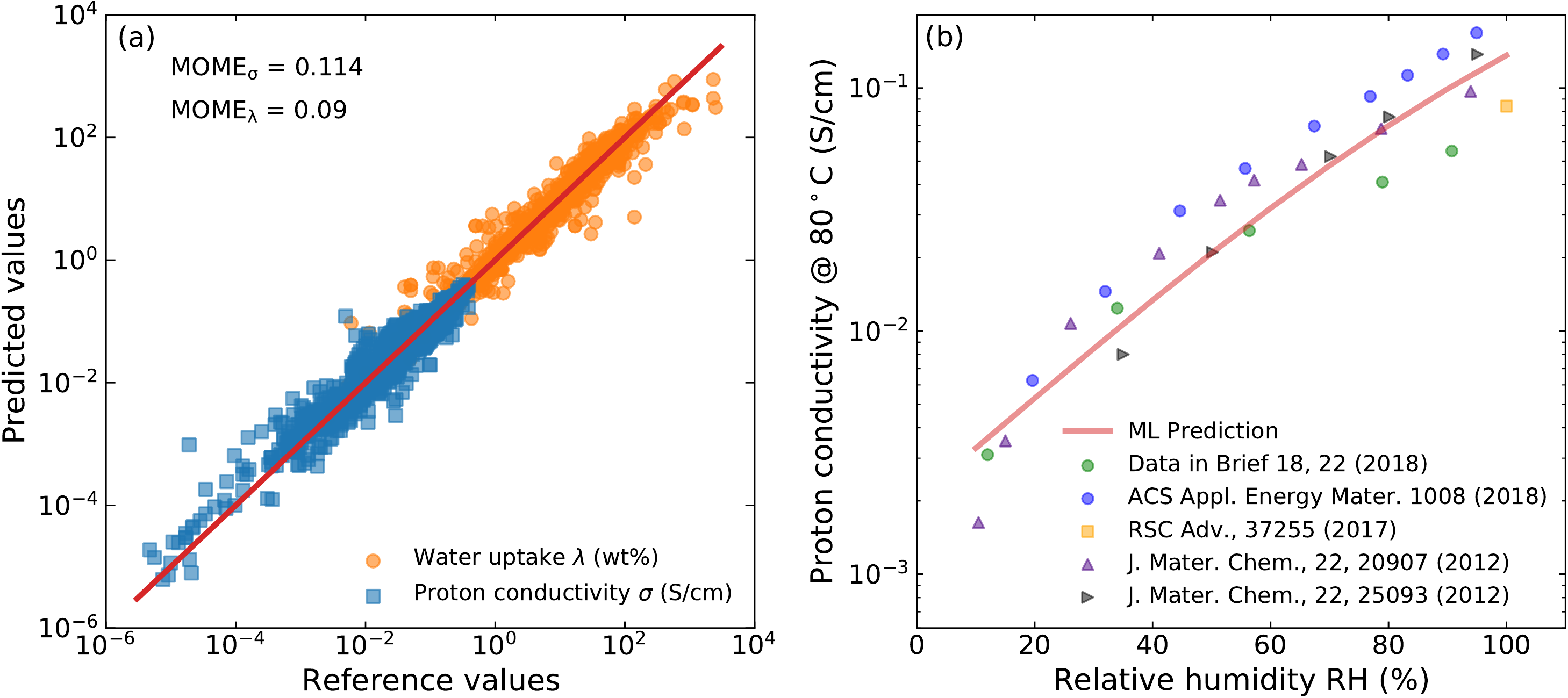}
	\caption{(a) The MT model M1, trained on two datasets of proton conductivity and water uptake, for predicting the proton conductivity $\sigma$, and (b) Proton conductivity $\sigma$, predicted by model M1 at $80^\circ$C as a function of the relative humidity RH, given in comparisons with experimental data. }\label{fig:models}
\end{figure}

Fig. \ref{fig:models} (a) visualizes the performance of the MT model M1 while Fig. \ref{fig:models} (b) shows the predictions of Nafion proton conductivity $\sigma$ predicted at $80^\circ$C as a function of the relative humidity. For M1, MOME $\simeq 0.1$ for both $\sigma$ and $\lambda$, which is about 2\% of the data range, indicating a good predictive performance. This conclusion is supported by Fig. \ref{fig:models} (b) which shows that compared to the available experimental data, M1 predicts both the values and the behavior of Nafion proton conductivity very well as a function of the relative humidity RH. Good predictions of the other models, i.e., M2, M3, M4, M5, and M6, for the key properties of Nafion were obtained and are given in Table \ref{table:model}. We note that Nafion properties such as $\mu_{\rm O_2}$, $\mu_{\rm H_2}$, and $E$ depend on the temperature $T$ and the relative humidity RH and were typically reported\cite{broka1997oxygen,mukaddam2016gas} as functions of these variables (see Table \ref{table:nafion}). Nevertheless, our training datasets do not have such information, thus the values of $\mu_{\rm O_2}$, $\mu_{\rm H_2}$, and $E$ predicted for Nafion also do not have $T$ and RH dependences. In particular, predicted $\mu_{\rm O_2}$, $\mu_{\rm H_2}$, and $E$ are close to the midpoints of reported data ranges, which were used as the thresholds of our screening.

\begin{table*}[t]
	\caption{Summary of six ML models, i.e., M1, M2, M3, M4, M5, and M6, all of them were trained using GPR, and their predictions for the key properties of Nafion. The error metric of M1 and M2 is MOME (dimensionless) while that of M3, M4, M5, and M6 is RMSE, given in the same unit with the respective property.}\label{table:model}

\begin{tabularx}{\textwidth}{@{} 
	>{\centering} p{1.05cm} 
	>{\RaggedRight} p{4.8cm}  
	>{\centering} p{1.2cm}  
	>{\centering} p{3.9cm} 
	>{\centering} p{1.3cm} 
	>{\centering} p{2.34cm} 
	>{\RaggedRight} p{0.1cm}
	}
\hline
\hline
	\multirow{2}{*}{Model} & \multirow{2}{*}{Key property} & \multirow{2}{*}{Unit} & \multirow{2}{*}{Dataset size} & Error& Prediction & \\
	& & & &metric& for Nafion & \\
\hline
	M1 &	Proton conductivity $\sigma$ &S/cm &2,137 ($\sigma$) + 1,879 ($\lambda$)&$0.114$& See Fig. \ref{fig:models} & \\
	M2 &	O$_2$ permeability $\mu_{\rm O_2}$ &Barrer &633 (in total of 2,624)&$0.082$&$10.7$&  \\
	M2 &	H$_2$ permeability $\mu_{\rm H_2}$ &Barrer &317 (in total of 2,624)&$0.057$&$37.2$ & \\
	M3 &	Electronic band gap $E_{\rm g}$ &eV &4,121&$0.25$& $7.90$ & \\
	M4 &    Glass trans. temp. $T_{\rm g}$ &K &8,513&$32.5$& $395.6$ & \\
	M5 &    Thermal decom. temp. $T_{\rm d}$ &K &4,649&$12.0$& $571.2$&  \\
	M6 &    Young's modulus $E$ & MPa &932&$256$& $155.7$& \\
\hline
\hline
\end{tabularx}
\end{table*}

\subsection{Identified candidates}
Our screening over 30,654 known polymers yields 48 candidates for PEM, 29 (19) of them were predicted to be better than Nafion in the liquid (vapor) water condition. In addition, 8 candidates for anode ionomer and 2 candidate for cathode ionomer were also identified. In fact, the requirement of having the sulfonate group (-SO$_3^-$) is a very strict condition, which significantly reduces the size of the search space from 30,624 polymers to 558 polymers (see Fig. \ref{fig:scheme}), and thus, limiting the number of candidates in our search. A full list of these candidates that includes all the available information, e.g., the SMILES strings, references, and predicted properties, can be found in Supporting Information. 

\begin{figure}[!ht]
\centering
\includegraphics[width=0.9\linewidth]{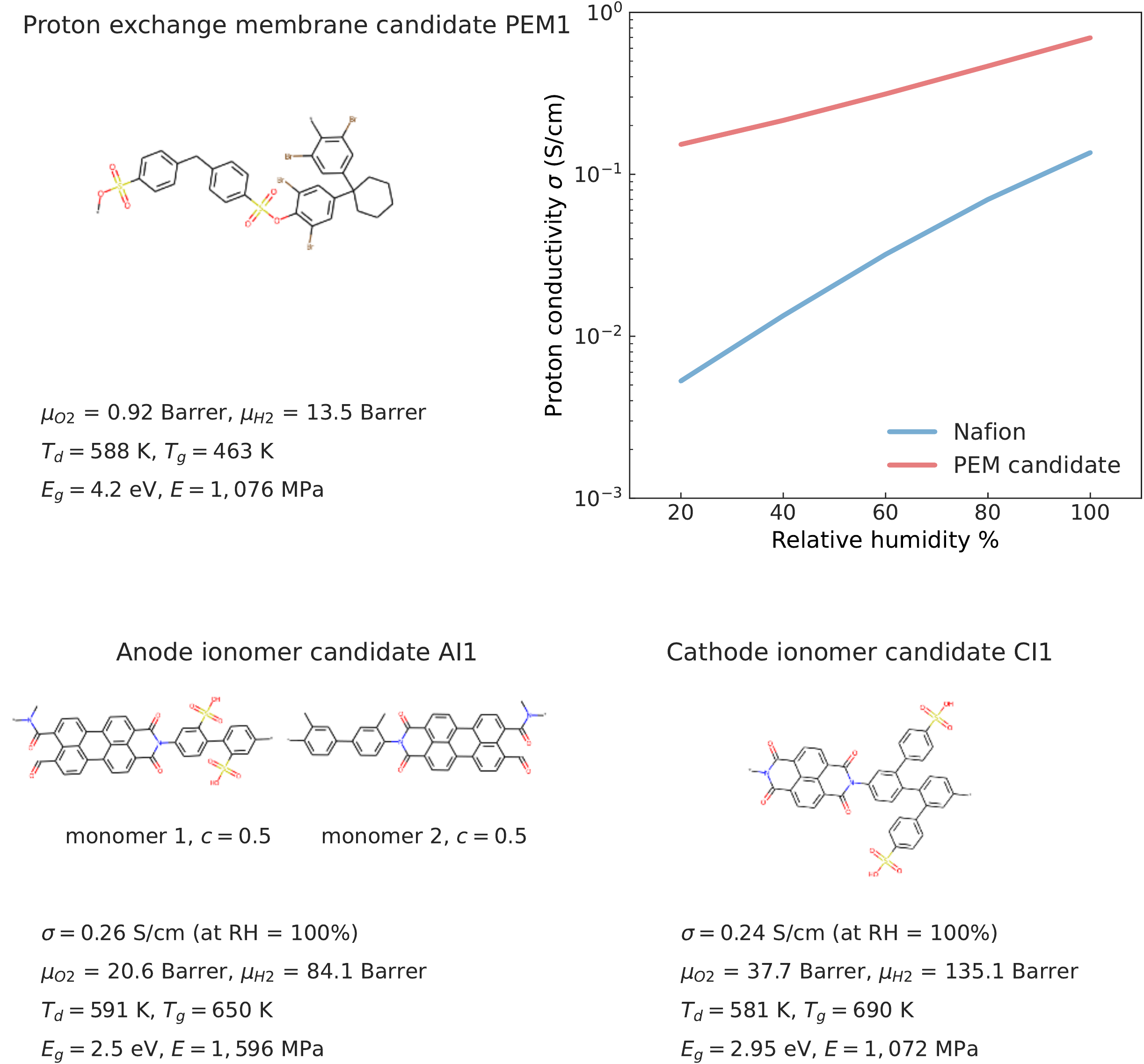}
\caption{Chemical structrure and key properties predicted for PEM1, AI1, and CI1, the representative candidate of PEM (top row), anode ionomer (lower left), and cathode ionomer (lower right), respectively.}\label{fig:cands}
\end{figure}

\begin{figure}[t]
\centering
\includegraphics[width=0.45\linewidth]{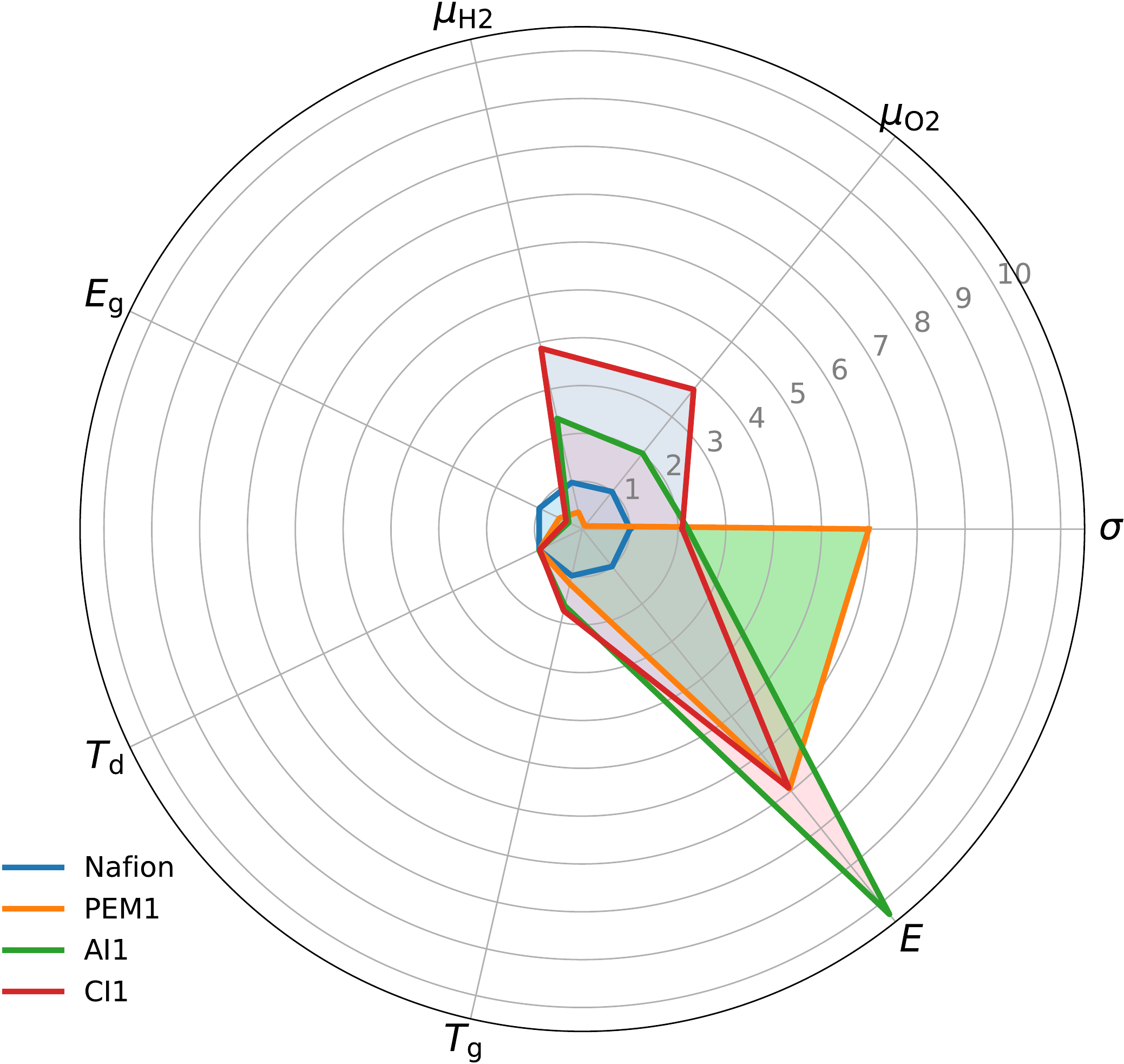}
	\caption{Property radar chart of PEM1, AI1, and CI1 whose details are given in Fig. \ref{fig:cands}. Their key properties are given in unit of the same properties of Nafion (scale is given in grey numbers). For PEM1, the proton conductivity $\sigma$ at RH=100\% is used.}\label{fig:cands_new}
\end{figure}

Three polymers identified to be candidates for proton exchange membrane, anode ionomer, and cathode ionomer and named PEM1, AI1, and CI2, are detailed in Fig. \ref{fig:cands} and their performance are visualized in Fig. \ref{fig:cands_new}. For each of them, the monomer chemical structure, monomer concentration $c$, and key properties predicted by the ML models developed in this work, are given. Because the PEM1 was predicted to have higher proton conductivity than Nafion in vapor water condition, $\sigma$ was given as a function of the relative humidity. On the other hand, the AI1 and CI1 were predicted at the liquid water condition, thus only the value of $\sigma$ at RH = 100\% is reported. The proton conductivity $\sigma$ predicted for these three polymers are about 2-3 times higher than that of Nafion. Specifically, very high values of Young's modulus $E$ suggest that all of them are mechanically very robust. We believe that PEM1, AI1, and CI1 could be good candidates for proton exchange membrane, anode ionomer, and cathode ionomer in fuel cell technology.

As the polymer search space, i.e., our candidate set, is composed of polymers previously synthesized and studied (but not necessarily for fuel cell applications), we hope that future experimental investigations of our recommended polymers for fuel cell applications will be possible in the near term. PEM1 is an aromatic (po1y)cycloaliphatic polysulfonate synthesized \cite{podgorski1987linear} from two monomers, i.e., 4,4'-(1-cyclohexylidene)-di-(2,6-dibromophenol) and 4,4'-diphenyldisulfonyl chloride, previously studied within a search for polymers with good physical, chemical and thermal properties. Therefore, some thermal and mechanical properties, including $T_{\rm d}$, of this polymer were measured \cite{podgorski1987linear}. Two experimental values of $T_{\rm d}$ reported for PEM1 are 583 K, which is the temperature of initial decomposition, and 613 K, which is the temperature of maximum velocity of decomposition from the thermal gravimetric analysis.\cite{podgorski1987linear} PEM1 is included in the $T_{\rm d}$ dataset we used to train M5, and our predicted value of $T_{\rm d} = 588$ K is very close to $598$ K, the averare of two reported values. Predictions on the other key properties, as shown in \ref{fig:cands}, suggest that this aromatic (po1y)cycloaliphatic polysulfonate is a good candidate for proton exchange membrane in fuel cells. Likewise, other candidates identified in this work have also been synthersized in some contexts, and they can now be reconsidered for specific applications in fuel cells.

The property radar chart shown in Fig. \ref{fig:cands_new} provides a visual overview of PEM1, AI1, and CI1 in terms of seven key properties identified for fuel cell applications. Because these properties are different by multiple orders of magnitude, the predicted properties of PEM1, AI1, and CI1 are given in unit of the same properties of Nafion. Compared to Nafion, all PEM1, AI1, and CI1 have much higher proton conductivity and Young's modulus. While PEM1 can better limit oxygen and hydrogen gases from penetrating and diffusing, AI1 and CI1 were predicted to promote the permeation of hydrogen and oxygen gas, respectively. The advantages of PEM1, AI1, and CI1 in these most important properties are obtained with small reductions of band gap $E_{\rm g}$, i.e., 4.5 eV for PEM1 and slightly less than 3 eV for AI1 and CI1. The tradeoff among these properties, as shown in Fig. \ref{fig:cands_new}, is an illustration for the multi-objective optimization problem in discovering new materials for specific applications.

\subsection{Chemical rules and multi-objective optimization}
Figs. \ref{fig:cands} and \ref{fig:cands_new} reveal the impressive proton conductivity predicted for PEM1, AI1, and CI1. We anticipate that the superior proton conductivity of these polymers is related to the sulfonate group -SO$_3^-$. In this work, we only consider those having this functional group partly because -SO$_3^-$ is hydrophilic and plays an essential role in capturing and holding water molecules in Nafion. However, the main reason we adopted this criteria is that our proton conductivity dataset is dominated by polymers involving the -SO$_3^-$ group, thus other chemistries are not well represented. We hope that when the proton conductivity dataset is further expanded and diversified, our model can handle other hydrophilic groups such as sulfonyl imide -SO$_2$N(H)SO$_2$CF$_3$,\cite{savett2002comparison,silva2019microhydration} -P(O)(OH)$_2$, and -P(O)(OMe)$_2$ that have also been suggested for fuel cell membranes.\cite{souzy2005proton} At this point, many more candidates for PEM, anode ionomer, and cathode ionomer can be found.

Fig. \ref{fig:cands_new} also shows that the Young's modulus $E$ of PEM1, AI1, and CI1 is substantially higher than that of Nafion. Such a mechanical strength, which is very useful for creating and maintaining a network of water channels in the materials, stems from the rigid framework of these polymers, which contain many benzene rings. In exchange, unsaturated bonds in these rings slightly reduce the band gap $E_{\rm g}$ of the identified polymers. In case of PEM1, a band gap of $\simeq 4.5$ eV is still good enough to make the polymer an electron insulator while securing the electrochemical stability of the PEM layer. In summary, Fig. \ref{fig:cands_new} offers a visual assessment to the multi-objective materials optimization problem that leads to the identification of PEM1, AI1, and CI1 as alternatives to Nafion for PEM, anode ionomer, and cathode ionomer, respectively.

\section{Discussion and conclusions}
Requirements placed on materials that can be used for a specific application can be translated into a set of desired physical/chemical properties. Within this work, which focuses on a search for alternatives of Nafion in fuel cell applications, we have established three sets of key properties needed for proton exchange membranes, anode ionomers, and cathode ionomers. Then, a series of predictive ML models were developed and used to evaluate the required properties of 30,654 polymers, ultimately identifying 60 polymers that can potentially be suitable for the targeted applications. In the next step, works to synthesize, test, and validate the identified polymers are desirable. 

The machine-learning and multi-objective driven strategy used in this work has its roots in polymer informatics, the subfield of polymer science that relies on generating, curating, and learning past data to quickly estimate the properties of previously unencountered polymers. This generic strategy has been demonstrated for discovering and designing polymer dielectrics in the past,\cite{Huan:design, Arun:design, Chiho:GA, Rohit:VAE,gurnani2021polyg2g, kamal2021novel,kamal2020computable} and, in principle, can be used for selecting materials in essentially any application (so long as screening criteria can be precisely stated and rapid predictive models are available for the properties in the screening criteria). We anticipate that multiple variants of this strategy can be developed and used in the near future.

\section*{Supporting Information}
The Supporting Information is available at \texttt{https://pubs.acs.org/doi/XXXX}. Visualized performance of 6 ML models (M1, M2, M3, M4, M5, and M6) developed in this work; Details on 48 candidates for proton exchange membrane, 10 candidates for anode ionomer, and 2 candidate for cathode ionomer; Predicted properties with uncertainties provided.

\begin{acknowledgement}
This work is financially supported by Toyota Research Institute through the Accelerated Materials Design and Discovery program. The authors thank XSEDE for computational support through Grant No. TG-DMR080058N and Christopher Kuenneth for technical help.
\end{acknowledgement}

\vspace{6mm}
\noindent \textbf{Notes}
\noindent Competing interests: The authors declare no competing financial interests.




\newpage

\section*{TOC graphic}
\begin{figure}[ht]
\centering
\includegraphics[width=8cm]{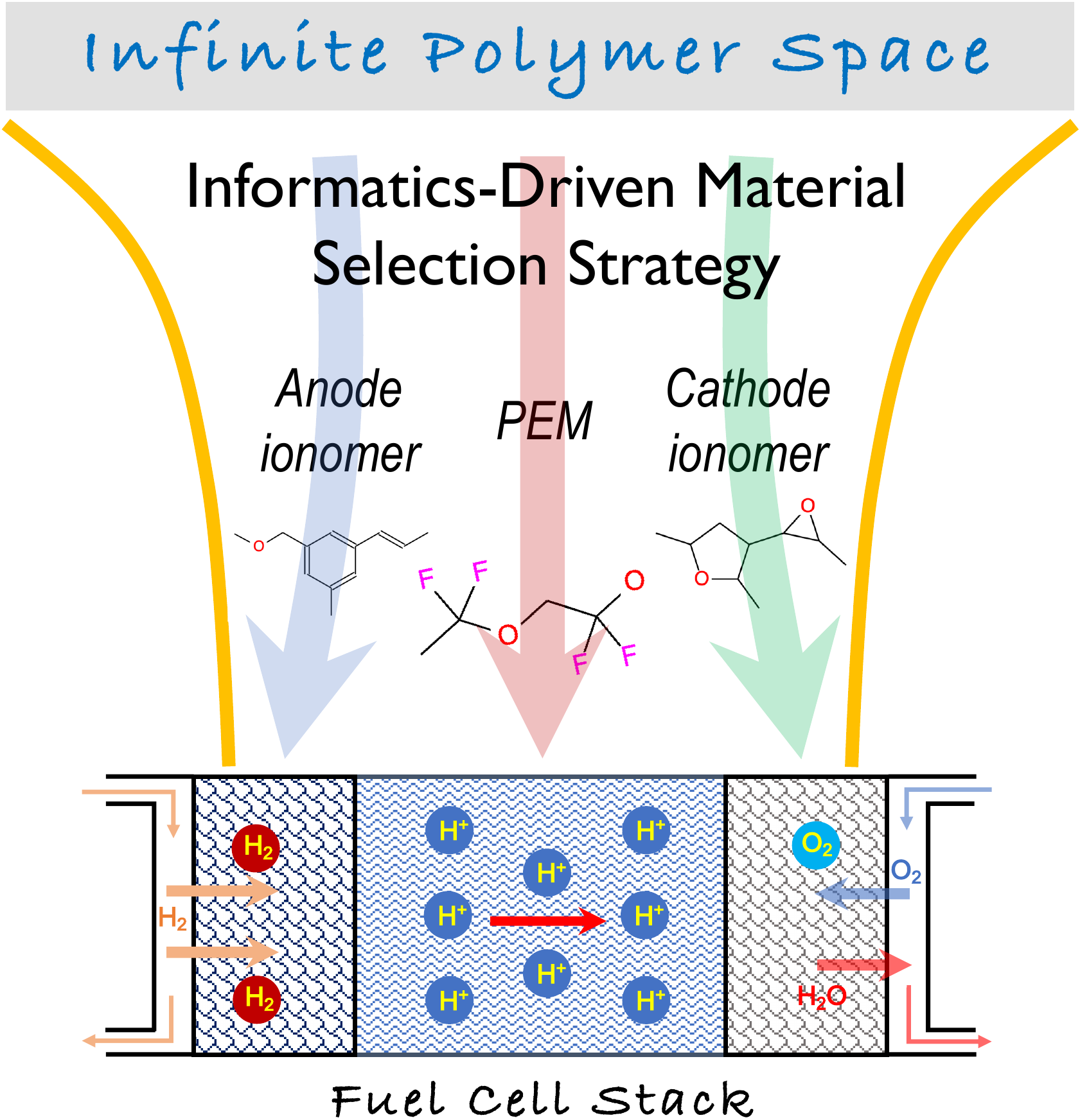}
\end{figure}

\newpage
\providecommand{\latin}[1]{#1}
\makeatletter
\providecommand{\doi}
  {\begingroup\let\do\@makeother\dospecials
  \catcode`\{=1 \catcode`\}=2 \doi@aux}
\providecommand{\doi@aux}[1]{\endgroup\texttt{#1}}
\makeatother
\providecommand*\mcitethebibliography{\thebibliography}
\csname @ifundefined\endcsname{endmcitethebibliography}
  {\let\endmcitethebibliography\endthebibliography}{}

\end{document}